\documentclass{vanvliet_paper}
\usepackage{enumitem}
\usepackage{mdframed}

\newcommand{\prevent}[2][0pt]{%
\marginnote{%
\begin{mdframed}[backgroundcolor=yellow!20,leftmargin=0cm,rightmargin=-0.5cm,innerleftmargin=1ex,innerrightmargin=1ex,innertopmargin=0pt,innerbottommargin=2ex]
\raggedright\small\singlespacing\textbf{Things that this guideline aims to\\prevent:}
\begin{itemize}[leftmargin=*]
    #2
\end{itemize}
\end{mdframed}
}[#1]
}

\newcommand{\fname}[1]{\href{https://github.com/AaltoImagingLanguage/conpy/blob/master/scripts/#1}{\code{#1}}}

\newacronym{MEG}{meg}{magnetoencephalography}
\newacronym{API}{api}{application programming interface}
\newacronym{GUI}{gui}{graphical user interface}
\newacronym{MRI}{mri}{magnetic resonance imaging}
\newacronym{fMRI}{\textnormal{f}mri}{functional magnetic resonance imaging}
\newacronym{ICA}{ica}{independant component analysis}
\newacronym{PCA}{pca}{principal component analysis}
\newacronym{EOG}{eog}{electrooculography}
\newacronym{CPU}{cpu}{central processing unit}
\newacronym{VCS}{vcs}{version control system}
\newacronym{MNE}{mne}{minimum norm estimates}

\title{Design guidelines for data analysis scripts}
\author[1]{Marijn van Vliet}
\affil[1]{Department of Neuroscience and Biomedical Engineering, Aalto University, Finland, marijn.vanvliet@aalto.fi}

\preprint
\begin{document}

\maketitle

\begin{abstract}

Unorganized heaps of analysis code are a growing liability as data analysis pipelines are getting longer and more complicated.
This is worrying, as neuroscience papers are getting retracted due to programmer error.
Furthermore, analysis code is increasingly published as the push towards open science continues, so the quality of your code becomes public knowledge.
In this paper, some guidelines are presented that help keep analysis code well organized, easy to understand and convenient to work with:
\begin{enumerate}
    \item Each analysis step is one script
    \item A script either processes a single recording, or aggregates across recordings, never both
    \item One master script to run the entire analysis
    \item Save all intermediate results
    \item Visualize all intermediate results
    \item Each parameter and filename is defined only once
    \item Distinguish files that are part of the official pipeline from other scripts
\end{enumerate}
In addition to discussing the reasoning behind each guideline, an example analysis pipeline is presented as a case study to see how each guideline translates into code. 


\end{abstract}

\section{Introduction}

The journey of the data from our measurement equipment to a figure in a publication is growing longer and more complicated.
new preprocessing steps have been developed to be added at the beginning of the pipeline\cite{Bigdely-Shamlo2015, Nolan2010, Jas2017}, new multivariate techniques that find a place in the middle\cite{Kriegeskorte2008, King2014, McIntosh2013}, and new statistical methods at the end\cite{Maris2007}.
Using these new techniques often requires writing pieces of programming code referred to as ``scripts'', and in accordance with the growing data analyses pipelines, these scripts also tend to increase in length and complexity.
When programming code becomes sufficiently convoluted, even the most experienced programmers will make mistakes, which can ultimately lead to erroneous conclusions.
It has happened that papers had to be retracted due to programmer error\cite{Casadevall2014, Miller2006, Merali2010}.
This paper is an effort to set some guidelines to manage the complexity of scripts.

In science, data analysis is performed at the cutting edge, where it is often inevitable that new pieces of programming code need to be written.
New methods are made available through software libraries first, being accessible through an \gls{API}, and only later as \gls{GUI} programs, if at all.
Furthermore, novel research ideas often require combining analysis techniques in new ways that are not possible with existing programs and hence require writing new code.
When data analysis pipelines grow, unorganized heaps of code become a liability.

Many fields, including neuroscience, are moving towards ``open science'', where data and analysis code are made public alongside the paper.
Researchers should be proud to share their analysis code, as it should be a testament to their scientific rigor and technological creativity.

An important aspect to organizing scientific code is to have the bulk of the analysis functionality implemented in the form of a software library that exposes a well designed \gls{API}\cite{Buitinck2013}.
The code can then be used in multiple scripts and user facing programs.
It is generally beneficial to use and extend an existing piece of software that is used by many, rather than developing a home-grown solution from scratch.
This is because the more people use a piece of software, the more likely it is that mistakes are spotted and corrected\cite{Eklund2016}.
There is a large body of literature on managing complexity and reducing the chance for programmer error in this context\cite{Beck2002, Hunt1999, Martin2008, McConnell2004, Wilson2014}.

The guidelines in this paper aim to translate some of this literature to the sub-domain of analysis scripts.
Scripts are pieces of code that use the functionality that is exposed by the \glspl{API} of software libraries to create data analysis pipelines according to the specific requirements of a single study.
This makes scripts somewhat of an outlier on the software landscape, as they are pieces of code that do not need to be reusable (pieces that need to be reusable are better implemented in a software library), only have to function correctly on one specific dataset (hence there is little need to test for ``edge cases'') and will generally only be used by yourself and your collaborators (save the occasional run for review purposes and replication studies).
Therefore, many of the standard practices of the software industry do not apply or need translation in order to arrive at concrete advice of what to do and what to avoid when writing analysis scripts.

Often, an analysis pipeline starts off as a simple script that runs a few operations and grows as more steps are added.
As the pipeline becomes more complicated, the overall organization and design of the pipeline must be occasionally re-evaluated, or it is likely to become convoluted and error prone.
The guidelines in this paper aim to facilitate a successful organization of the analysis code, thereby keeping the complexity of data analysis scripts within tolerable limits, capitalize on the advantages of scripting and offset the disadvantages.

This article is a guide for those who have already written their own data analysis pipelines and wish to improve their designs.
Those looking for information on how to get started writing a data analysis pipeline for neuroimaging are referred to other works\cite{Jas2018, Popov2018, Tadel2019, Andersen2018, Andersen2018a}.

To move beyond mere truisms, the analysis pipeline developed by \textcite{VanVliet2018a} has been extended to implement all guidelines in this paper, and will be used as case study. In the case study, the practical consequences are discussed of each guideline in terms of code, which can serve as an example when implementing your own analysis pipelines.

The example pipeline starts from the raw \gls{MEG} and structural \gls{MRI} data from the \textcite{Wakeman2015} faces dataset and performs several artifact reduction steps, source estimation, functional connectivity analysis, cluster permutation statistics and various visualizations.
The size and complexity of the pipeline is representative of that of the pipelines in modern studies at the time of writing.
Where \textcite{VanVliet2018a} gives a detailed explanation of all analysis steps, the current paper focuses on the design decisions that were made during the implementation.
You can find the code repository for the analysis pipeline at: \url{https://github.com/aaltoimaginglanguage/conpy}.
Of special interest is the \fname{scripts} folder of that repository, which contains the analysis code itself.

The ``Application of the guideline to the example analysis'' sections refer frequently to the example code and it is recommended to study these sections and the code side by side. The electronic version of this document contains many hyperlinks to sections of the code, which the reader is encouraged to follow to see how the guidelines can be implemented in practice.
Hyperlinks are typeset in dark blue.

By nature, analysis scripts are specific to a single study.
Hence, the primary intention for the example pipeline is to be a source of ideas to use when writing your own analysis pipelines.
However, should you wish to construct a pipeline similar to the example, a stripped down version can be found at \url{https://github.com/aaltoimaginglanguage/study_template} along with instructions on how to use it as a template for new analysis pipelines.

\section{Guidelines}

\subsection{Guideline 1: Each analysis step is one script}

An effective strategy to reduce software complexity is to break up a large system into smaller parts\cite{Hofmann2004, Parnas1972}.
The first guideline is therefore to isolate each single step of an analysis pipeline into its own self-contained script.
This greatly reduces complexity by allowing us to reason about the pipeline on two levels.
At the lower level, we can reason about the implementation of a single step, while ignoring the rest of the pipeline for a moment.
At the higher level, we can treat the individual steps as ``black boxes'' and focus on how they are combined together to form the complete pipeline (see also guideline 3), ignoring their implementation for a moment.

\prevent[-2.5cm]{
    \item script becomes ``spaghetti code''
    \item excessive running time of the script
    \item parts of the script are commented out in order to skip a time-consuming step
    \item if-statements being used to toggle parts of the script on and off
}

This raises the question of what exactly constitutes a single analysis step.
The decision of where to ``cut'' the pipeline can be made from different perspectives: by complexity, by theme, and by running time.

\subsubsection{Complexity perspective}
The purpose of the guideline is that each individual script should be easy to understand and reason about, so one way to define a single step is by its complexity: if a single script becomes too complex to be easily understood as a whole, it should be split up into smaller steps if possible.

\subsubsection{Thematic perspective}
Ideally, understanding one script should not require knowledge of another script.
If each script can be viewed as a self-contained box that performs a single task, the pipeline as a whole becomes simply a collection of these boxes that are executed in a specific order.
A script should therefore aim to implement a single task, not multiple, and implement it completely, not only part of it.

\subsubsection{Time perspective}
While running the entire analysis pipeline may take days, a single script should finish in a reasonable time.
This invites frequent testing as you iteratively develop the script, allows you to quickly evaluate the effect of a parameter, and also makes it painless to ensure that the latest version of the script matches the latest result.
When the running time grows to the point where you are tempted to continue working on the script while a run is still in progress, the script should be split into smaller steps if possible.

These perspectives may clash with one another, so compromises are sometimes necessary.
We will now look how these perspectives influenced the way the \textcite{VanVliet2018a} pipeline was split into individual steps.

\subsubsection{Application of the guideline to the example analysis}

\begin{figure}
    \includegraphics[width=\textwidth]{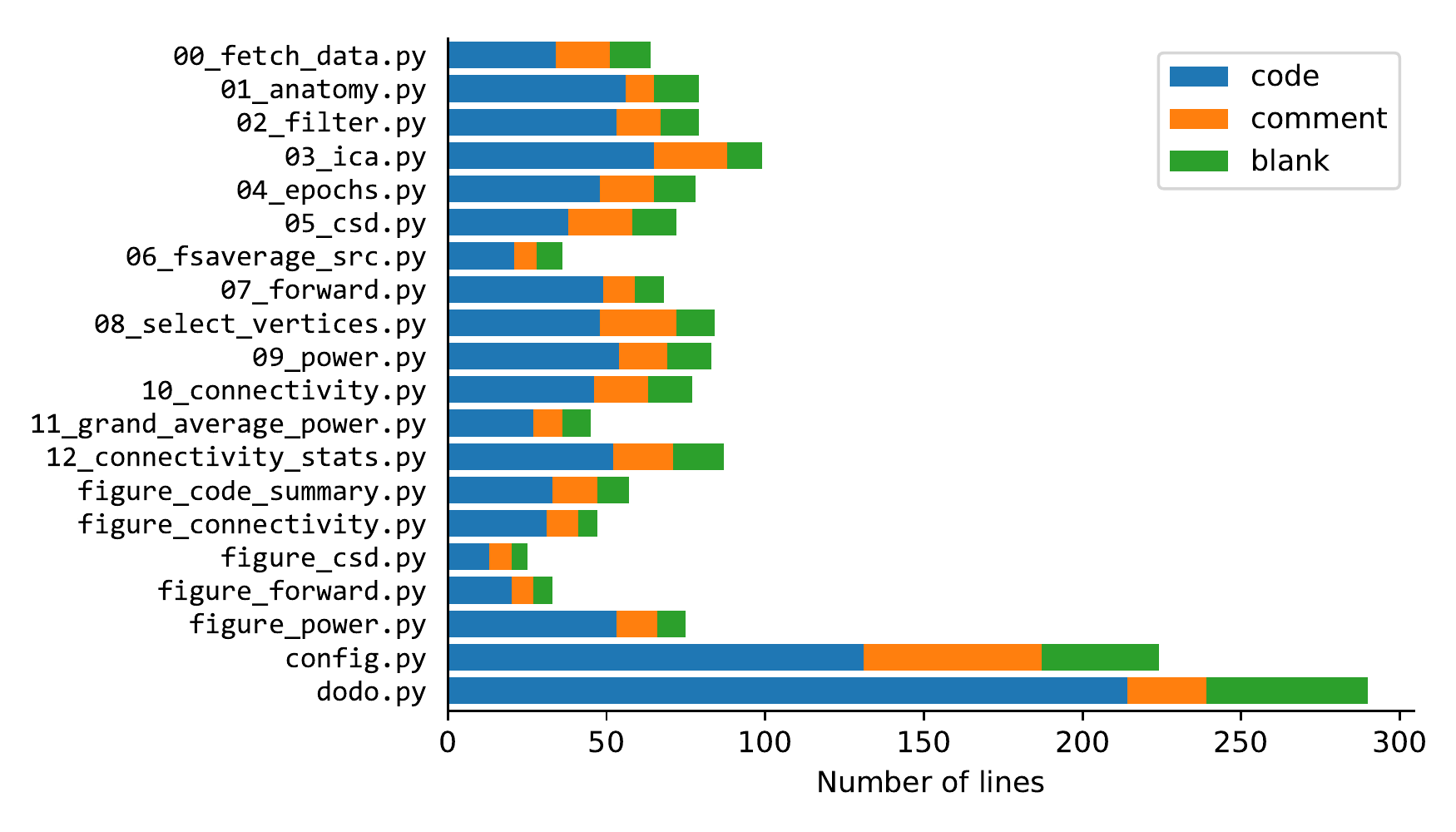}
    \caption{For each script in the analysis pipeline, the number of lines of the file, broken down into lines of programming code (code), lines of descriptive comments (comment) and blank lines (blank). The first 13 scripts perform data analysis steps, the next 5 scripts generate figures, the \fname{config.py} script contains all configuration parameters and the \fname{dodo.py} script is the master script that runs all analysis steps on all recordings.}\label{fig:summary}
\end{figure}

The \textcite{VanVliet2018a} pipeline consists of 13 analysis scripts that process the data, 5 visualization scripts that construct the figures used in publications, a configuration file (\fname{config.py}) and a ``master'' script that calls the individual analysis scripts (\fname{dodo.py}, see guideline 3).
Each of the 13 analysis scripts implements a single step in the analysis and are numbered to indicate the sequence in which they are designed to be run.
While the scripts need to be run in sequence once, they can be run independently afterwards.
The analysis scripts are relatively short (\autoref{fig:summary}), containing an average of 40.8 lines of code (std. 14.8), while the configuration (see guideline 6) and master scripts (see guideline 3) are longer.

Often, the reasoning behind the scope of each script was made from a thematic perspective.
For example, one script performs the source estimation (\fname{09_power.py}) and another the connectivity estimation (\fname{10_connectivity.py}).
However, the decision to split the artifact reduction steps into two scripts (\fname{02_filter.py}, \fname{03_ica.py}) was made from a time perspective.
Since the \gls{ICA} computation takes time, it was split off into its own script to avoid having to repeat it unnecessarily.
Finally, the decision to split up the construction of the forward models (i.e.~leadfield) into three steps (\fname{06_fsaverage_src.py}, \fname{07_forward.py}, \fname{08_select_vertices.py}) was made from a complexity perspective.

\newpage
\subsection{Guideline 2: A script either processes a single recording or aggregates across recordings, never both}

\prevent{
    \item excessive running time of the script
    \item multiple versions of the Big Loop that operate on different sets of subjects, all but one commented out
    \item a copy/paste of the Big Loop for each analysis script
}

A compelling reason for performing data analysis using scripts instead of, for example, using a \gls{GUI}, is the ease of repeating (parts of) the analysis.
Every time the scripts are run, the computer will perform exactly the same tasks in exactly the same order, eliminating the possibility of mix-ups in this regard.
This allows you, for example, to efficiently test the effect of changing a single parameter, while ensuring all subsequent analysis steps remain the same.
To capitalize on this advantage, analysis scripts should be organized such that it is easy to run only selected parts of it, without having to modify (e.g. ``commenting out'') the code itself.

In neuroscience, it is common to apply the same data processing steps to multiple recordings.
For example, a frequently seen construct is the Big Loop over data from multiple participants.
The second guideline states a separation of duties: a script is either a processing script, or an aggregation script.
Processing scripts perform data analysis only on a single recording, passed as a parameter from the command line, and do not have the Big Loop (the script is applied to all recordings in a separate ``master'' script (see guideline 3)).
Aggregation scripts have the Big Loop to collect the processed data from multiple recordings, with the sole purpose of aligning the data (e.g.~morphing to a template brain) and computing an aggregate (e.g.~a grand average or statistics).

This reduces the complexity of the code, since it allows the reader to either focus on the intricacies of a data processing step, without having to worry about how the data is later reconciled across recordings, or to focus solely on the details of how multiple datasets are aligned and combined.

Following the guideline also makes the development process more efficient.
It allows for a smooth workflow for the common scenario in which the script is tested on one subject during development, then an attempt is made to run it on all subjects using the master script, problems are found that only arise for certain subjects, and finally the script is re-run once more on all subjects.

\subsubsection{Application of the guideline to the example analysis}

In the example pipeline, there is a strict separation between scripts that perform data analysis on a single participant (steps 0--10) and scripts that aggregate across participants (steps 11 and 12).
The scripts implementing steps 0--10 all take a \href{https://github.com/AaltoImagingLanguage/conpy/blob/2f9926334fecc57de0e38b8f5124385dc661f17f/scripts/02_filter.py#L14-L15}{single command line parameter} indicating the participant to process.
This is implemented with the \href{https://docs.python.org/3/library/argparse.html}{\code{argparse}} module of the standard Python library, which facilitates the generation of a helpful error message when this parameter is omitted, along with documentation on how to run the script.
Not only does this help to keep the number of lines of code and the running time of the script down (\autoref{fig:summary}), it also opens up the possibility for the \href{https://github.com/AaltoImagingLanguage/conpy/tree/master/scripts/dodo.py}{master script} (see guideline 3) to automatically skip running the script for participants that have already been processed earlier.

\subsection{Guideline 3: One master script to run the entire analysis}

Once the individual steps have been implemented as a collection of scripts, the pipeline can be assembled in a ``master'' script that runs all the steps on all the data.
This master script is the entry point for running the entire analysis and therefore the third guideline states that there should ideally only be one such script.

\prevent{
    \item excessive complexity
    \item confusion as to which order the scripts should be run in
    \item having to manually run several scripts in order to complete the analysis and forgetting to re-run one
    \item scripts are changed, but not re-run, causing the result to be out of sync with the code 
    \item excessive running time of the analysis pipeline when only a single step has changed
    \item a copy/paste of the Big Loop for each analysis script
}

By having a strict separation between the scripts that implement the individual steps and the master script, it becomes possible to view the pipeline on two levels: the implementation details of each single step, and how the steps fit together to build the pipeline.
To understand the latter, the master script provides the ``floor plan'' of the analysis, which can be studied without having to go into detail on how each individual step is performed.
Hence, the only function that the master script should perform is to call the other scripts in the correct order.
Actual data analysis steps, including logic for combining results across scripts, should always be performed in a separate script, which is in turn called from the master script.

Speed is a very important aspect of an analysis pipeline, as it encourages practices that reduce the likelihood of errors.
Speed encourages incremental development, running the pipeline often during development to check the intermediate results.
Speed encourages exploration, trying different parameters and approaches to obtain the best result possible.
And last but not least, speed encourages re-running a script every time it has changed.

Apart from using efficient algorithms, the key to obtaining speed is to never repeat a time-consuming calculation unnecessarily.
If all analysis steps are properly isolated from one another (see guidelines 1 and 2) and all intermediate results are properly stored (see guideline 3), analysis steps for which the corresponding scripts have not changed, need not be run again if a script further down the pipeline has changed.

A compelling advantage of scripting is that the code serves as a complete transcript of exactly what analysis steps were performed.
However, this transcript is only correct if the latest version of the code is also the version that was used to produce the latest results.
During the development of the scripts, we commonly make changes, re-run the script, inspect the result, and make more changes.
If we are not careful, the code and results may become desynchronized, especially when multiple versions of the code and results are in play simultaneously.
Putting misplaced trust in a wrong transcript can be very frustrating when attempting to reproduce a result.

Keeping track of which scripts have changed since they were last ran, and which scripts consume the output produced by which other scripts (known as the dependency graph, see figure \autoref{fig:graph}), is a task that has been studied in great detail in the area of software engineering and many specialized tools, known as ``build systems''\footnote{For example, here are some build systems that are optimized for  creating data analysis pipelines:\\
\url{https://snakemake.readthedocs.io}\\
\url{https://pydoit.org}\\
\url{https://luigi.readthedocs.io}
}, are available to perform the required bookkeeping tasks.
Writing the master script using a build system will allow fine grained control over which steps to run on which recordings, while skipping steps that are ``up to date''.

\subsubsection{Application of the guideline to the example analysis}

\begin{figure}
    \includegraphics[width=\textwidth]{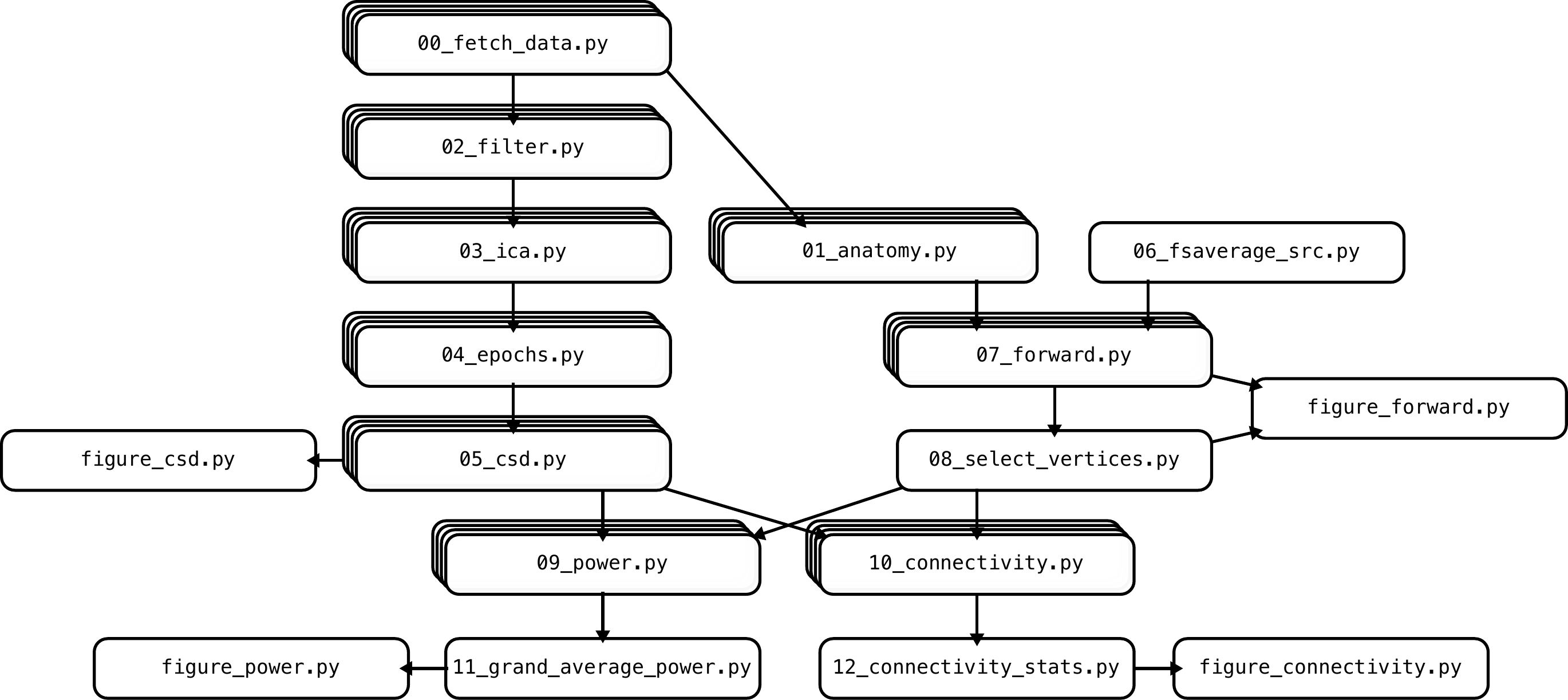}
    \caption{Dependency graph showing how the output of one script is used by another. Stacked boxes indicate scripts that are run for each participant.}\label{fig:graph}
\end{figure}

The master script of the example pipeline, \fname{dodo.py}, is implemented using the \href{https://pydoit.py}{\code{pydoit}}\footnote{\url{https://pydoit.org}} build system.
In the script, all analysis steps are described as \href{https://github.com/AaltoImagingLanguage/conpy/blob/2f9926334fecc57de0e38b8f5124385dc661f17f/scripts/dodo.py#L18}{``tasks''}, which steps 0--10 having a \href{https://github.com/AaltoImagingLanguage/conpy/blob/2f9926334fecc57de0e38b8f5124385dc661f17f/scripts/dodo.py#L28-L38}{``subtask'' for each participant}.
Each task is associated with one of the analysis scripts, along with a list of files the script uses and produces.
This allows the build system to work out the dependency graph of the analysis pipeline (\autoref{fig:graph}).

The build system keeps track of which tasks are ``up to date'', meaning the latest version of the analysis scripts of all analysis steps up to and including the current step have all been run.
This means that the entire analysis pipeline can be run often and cheaply: all steps that are up to date will be skipped.
Making a change anywhere within the analysis code will prompt the recomputation of all the steps that need to be re-run.

The build system also provides a \href{http://pydoit.org/tutorial_1.html#doit-command-line}{set of commands} that allow for executing specific parts of the pipeline, for example, a single analysis step on all participants, a few specific steps on a few specific participants, etc., without having to change (e.g.~comment out) any code.

\subsection{Guideline 4: Save all intermediate results}

\prevent{
    \item variables being manipulated across multiple scripts
    \item debugging a script taking a long time due to having to re-compute everything from scratch every time the script runs
    \item erroneous output is generated in the middle of the script, but subsequent processing makes the result appear reasonable at the end of the script
}

First, from the complexity perspective, it is important that each script can function in isolation and does not rely on data that was left in memory by another script.
The more each script can be isolated from the rest of the pipeline, the easier it is to understand and represent as a self contained black box.

The fourth guideline states that all intermediate results generated during the execution of a script should be saved to disk, if feasible.
Having snapshots of the data as it passes through the pipeline has numerous advantages.

Another big advantage is that it makes it possible to skip any data processing steps that are unchanged since the last time the pipeline was executed (see also guideline 3).
This makes it possible to re-run small portions of the pipeline quickly, for example to debug a problem, or to assess the effect of some parameter.

Finally, having all intermediate results readily available facilitates manual checks and exploration of the data.
The ability to jump into an interactive session and quickly load the state of the data at any desired location in the analysis pipeline is an effective way to verify that a script produced the intended result.

In cases of limited storage capability, the cost of storage need to be weighed against the benefits of not having to recompute the result.
Here, it helps to identify computational bottlenecks and only store the minimum amount of data needed to bypass the bottleneck when re-running the script.
For example, for linear data transformations such as \gls{PCA}, \gls{ICA}, \gls{MNE} and beamformers, the computational bottleneck is in the computation of the transformation matrix, which by itself does not take much disk space.
Applying the transformation matrix to the data is computationally cheap, while the result may take up a large amount of disk space.

\subsubsection{Application of the guideline to the example analysis}

In the example analysis pipeline, each script begins by \href{https://github.com/AaltoImagingLanguage/conpy/blob/2f9926334fecc57de0e38b8f5124385dc661f17f/scripts/05_csd.py#L25-L27}{loading the data} that were produced by previous analysis steps as requires.
Each script ends by \href{https://github.com/AaltoImagingLanguage/conpy/blob/2f9926334fecc57de0e38b8f5124385dc661f17f/scripts/05_csd.py#L50}{saving all data} that was produced by the script.
This includes the processed \gls{MEG} data, but also, for example, the \gls{ICA} decomposition matrix, along with the indices of the \gls{ICA} components that were judged to correspond to eye-blink contaminants.

\subsection{Guideline 5: Visualize all intermediate results}

\prevent{
    \item researcher is operating ``in the blind''
    \item erroneous output is generated in the middle of the script, but subsequent processing makes the result appear reasonable at the end of the script
    \item result figures no longer match the data files after a script has been re-run
}

Data analysis pipelines, such as those used in neuroscience, are sufficiently complex that failures should be expected and planned for.
When designing the pipeline, think about the system for catching errors when they happen.

While using analysis scripts instead of a instead of a \gls{GUI} offers many advantages, a severe disadvantage is the lack of direct feedback.
Since the result is usually not immediately visualized, errors may stay hidden for a long time.
Programming a computer is not unlike receiving a wish from a mischievous genie: you will get exactly what you asked for, but not necessarily what you wanted.
As long as the final result of a series of processing steps looks reasonable, intermediate steps might contain nonsensical results that we would never know about unless we take care to check everything.
Therefore, an analysis pipeline should invite frequent visual checks on all intermediate results.

The fifth guideline states that for each intermediate result, the script should create a visualization of the result and save it to disk.
This does not need to be a publication ready figure, but must provide a visual confirmation that the data analysis operation had the intended result.
By re-creating the figures every time a script is run and overwriting the file on disk, the figure remains up to date.

After running all the analysis scripts, a complete visual record should be available of the data as it moves through the pipeline.
Care should be taken that the order of the figures matches that of the analysis steps.
Such a record invites frequent visual checks of the obtained results and therefore somewhat offsets the main advantage that \gls{GUI} programs have over scripting.

\subsubsection{Application of the guideline to the example analysis}

Whenever an intermediate result is saved to disk in the example analysis pipeline, a \href{https://github.com/AaltoImagingLanguage/conpy/blob/2f9926334fecc57de0e38b8f5124385dc661f17f/scripts/03_ica.py#L82}{simple visualization} is also created and \href{https://github.com/AaltoImagingLanguage/conpy/blob/2f9926334fecc57de0e38b8f5124385dc661f17f/scripts/03_ica.py#L83-L96}{added to a ``report'' file}.
The main analysis package used in the example pipeline, MNE-Python\cite{Gramfort2013}, provides a \code{Report} class that
compiles a set of figures into a single HTML file.
Each script adds (and overwrites) figures to the same report, which will grow in length as more scripts are run.
The resulting HTML file contains an easy to navigate visual record of the data flowing through the pipeline.
Each participant has their \href{https://users.aalto.fi/~vanvlm1/conpy/reports/}{own report file}.

\subsection{Guideline 6: Each parameter and filename is defined only once}

It is not uncommon that a parameter is used in multiple scripts.
The sixth guideline states that the value of each parameter should be defined in one place.
Instead of copying the value of a parameter into all scripts that need it, the parameter should be imported, i.e., the programmer specifies the location where the parameter is defined and the programming language will take care of fetching the value when it is needed.
In the programming literature, this is referred to as the "don't repeat yourself" (DRY) principle\cite{Martin2008}.

\prevent{
    \item the same parameter, defined at two locations, with two conflicting values
    \item when changing a parameter, not knowing where else in the code the same change should be made
    \item wasting time copy/pasting things across multiple scripts
}
Importing, rather than copying, eliminates a common source of errors.
When we change the value of a parameter, we may not be aware that we need to change it in multiple locations (either we forgot about the copies or we didn't know about them in the first place), resulting in different values being used at different locations and hence errors that can be very difficult to spot as long as the final result looks reasonable.

A good tactic for managing parameters is to create a single configuration script, which sole function is to define the values for all parameters.
It makes it obvious were to look for the definition of a parameter and decreases the chances of accidentally defining the same parameter at two different locations.

Filenames are also parameters, and ones that are commonly shared across scripts too: one script producing a file that another script uses.
Just like other parameters, the guideline mandates that all filenames should be defined once and imported (not copied) by scripts that need it.
It is not uncommon for a filename to change when parts of the analysis pipeline are added or removed.
By ensuring the change needs to be made in only one location, scripts are less likely to load data from the wrong file.

A good tactic for managing filenames is to define templates for them in the configuration file.
The templates can have placeholders for things like the participant number or experimental condition.
See the implementation example for a more thorough explanation.

\subsubsection{Application of the guideline to the example analysis}

The example pipeline has a central configuration script \fname{config.py} which defines all relevant parameters for the analysis, such as \href{https://github.com/AaltoImagingLanguage/conpy/blob/2f9926334fecc57de0e38b8f5124385dc661f17f/scripts/config.py#L56-L57}{filter settings}, \href{https://github.com/AaltoImagingLanguage/conpy/blob/2f9926334fecc57de0e38b8f5124385dc661f17f/scripts/config.py#L114-L140}{the list of subjects}, \href{https://github.com/AaltoImagingLanguage/conpy/blob/2f9926334fecc57de0e38b8f5124385dc661f17f/scripts/config.py#L142-L153}{the experimental conditions}, and so forth.
All analysis scripts \href{https://github.com/AaltoImagingLanguage/conpy/blob/2f9926334fecc57de0e38b8f5124385dc661f17f/scripts/02_filter.py#L7-L8}{import the configuration file} and thereby gain access to the parameters.
Whenever a parameter needs to be changed or added, the configuration file is the single authoritative location where the edit needs to be made and the change is propagated to all analysis scripts.

The \fname{config.py} script starts by offering some \href{https://github.com/AaltoImagingLanguage/conpy/blob/2f9926334fecc57de0e38b8f5124385dc661f17f/scripts/config.py#L25-L47}{machine specific parameters}, such as the number of \acrshort{CPU}-cores to dedicate to the analysis and where on the disk the data is to be stored. The configuration script \href{https://github.com/AaltoImagingLanguage/conpy/blob/2f9926334fecc57de0e38b8f5124385dc661f17f/scripts/config.py#L22-L23}{queries the hostname}, so that different parameters can be specified for different machines.

Since the pipeline stores all intermediate results and their visualizations, there is a large number of filenames to deal with. In many cases, each filename is used four times: \href{https://github.com/AaltoImagingLanguage/conpy/blob/2f9926334fecc57de0e38b8f5124385dc661f17f/scripts/07_forward.py#L49}{once in the script generating the file}, \href{https://github.com/AaltoImagingLanguage/conpy/blob/2f9926334fecc57de0e38b8f5124385dc661f17f/scripts/08_select_vertices.py#L24}{once in the script using  the file}, and \href{https://github.com/AaltoImagingLanguage/conpy/blob/2f9926334fecc57de0e38b8f5124385dc661f17f/scripts/dodo.py#L136-L164}{twice in the master script} \fname{dodo.py}.
For this reason, a helper class \href{https://github.com/AaltoImagingLanguage/conpy/blob/master/scripts/fnames.py}{\code{Filenames}} has been written that offers an efficient way to manage them.
The class is used to create an \href{https://github.com/AaltoImagingLanguage/conpy/blob/2f9926334fecc57de0e38b8f5124385dc661f17f/scripts/config.py#L162}{\code{fname} namespace} that contains \href{https://github.com/AaltoImagingLanguage/conpy/blob/2f9926334fecc57de0e38b8f5124385dc661f17f/scripts/config.py#L164-L221}{short aliases for all filenames} used throughout the pipeline.
It also leverages Python's \href{https://docs.python.org/3/library/string.html#formatstrings}{native string formatting language} to allow quick generation of lists of filenames that adhere to a pattern (e.g., \code{"sub01_raw"}, \code{"sub02_raw"}, \ldots).


\newpage
\subsection{Guideline 7: Distinguish files that are part of the official pipeline from other scripts}

\prevent{
    \item inability to distinguish incomplete or flawed scripts from proper ones
    \item not knowing what files are relevant to the main analysis
    \item not knowing what script is the master script that will run the entire analysis
    \item not knowing which version of the script was last run on the data
}

The development of a complex analysis pipeline is seldom a straightforward path from start to finish.
Rather, ideas get tried and discarded, mistakes are made and fixed, and smaller analysis are made on the side.
This all causes a tendency for scattering miscellaneous files around, littering the folders that contain the main pipeline, obfuscating what is relevant and what is not.
However, since creative freedom is important, it would be counterproductive to strive for a perfectly clean workspace all the time.
Rather, the occasional mess should be embraced, as well as the resulting responsibility to clean up after ourselves.

The seventh guideline calls for an organization system that distinguishes between files that are in a stable state and part of the main pipeline, and files that are work in progress, temporary, or part of analyses on the side.
A good system reduces the effort of the cleaning process, making it easier to commit to a regular tidying up of the virtual workplace.
It is important that the system does not become burdensome, as a simple system that is actually used is better than a more powerful one that is not.

This can be implemented in whatever way suits your workflow best, ranging from simply maintaining a rigid naming convention and folder structure, to using more powerful tools such as a \gls{VCS}.
Note that a \gls{VCS} is not an organization system in itself, but merely a tool for implementing one.
It is up to the data analyst to devise their own system and have the discipline to stick to it.

\subsubsection{Application of the guideline to the example analysis}

During the development of the pipeline, many scripts were written to try out different analysis approaches, conduct tests and do miscellaneous other tasks.
The naming system was such that all analysis scripts that are officially part of the pipeline are either prefixed with a number (\code{00_}--\code{12_}) and all scripts that produce figures for the manuscript with \code{figure_}.
From time to time, any script lacking such a prefix would be closely scrutinized to determine whether is was still relevant, and if not, deleted.

Deleted files were never truly gone though, as the project is managed by the \gls{VCS} ``Git''\footnote{\url{https://git-scm.com}}\cite{Chacon2019}.
Git keeps track of the history of a file, allowing to return to previous versions, as well as parallel copies when doing something experimental.
Although \glspl{VCS} are primarily used to facilitate collaboration on a software project, they are useful even when working alone\cite{Vuorre2018}.
For one, they provide a crucial backup service, allowing recovery from mistakes and thus freedom to experiment and making bold decisions.
Secondly, although \glspl{VCS} do not impose any organizational structure for managing multiple copies of files, they facilitate creating and maintaining one.

\section{Conclusion}

Following the given guidelines improves the chances of the analysis code being correct, by aiming for code that is easy to understand.
The key to reducing the complexity of data analysis scripts is to cut up the pipeline into bite-sized chunks.
Therefore, the guidelines state that each step of the pipeline should be implemented as a separate script that finishes in a reasonable time, has little to no dependencies on other scripts, writes all intermediate results to disk and visualizes them.
In addition, one master script should exist that calls the other scripts in the correct order to execute the complete analysis pipeline.

Writing understandable code is a skill that can be honed by forming good habits.
Whenever a problem arises in a pipeline, there is an opportunity to look beyond the specific problem to the circumstances that allowed the problem to occur in the first place and the formation of new habits to prevent such circumstances in the future.
However, it is important not to become bogged down in rules.
Every new project is a chance for reviewing your habits: keep things that were beneficial and drop things that were not or which costs exceed their utility.

Be aware that there is a lot of software tooling available to automate repetitive tasks and perform bookkeeping.
Whenever a rule needs enforcing or a repetitive action needs performing, there is likely a software tool available to automate it.
However, while they can make it easier to adopt good habits and keep the code organized, they cannot do the job by themselves.
Ultimately, it is up to the data analyst to keep things tidy and re-evaluate the design of the analysis pipeline as it grows.
When and how to do this is best learned through experience.
By reflecting at the end of each project what were good and bad design choices, the analysis pipeline of the next project will be better than the last.

\section{Acknowledgements}

Thanks goes out to all the members of the department of Neuroscience and Biomedical Engineering (NBE) at Aalto University who contributed to the discussion during the lab meeting concerning data analysis pipelines.
MvV is supported by the Academy of Finland (grant 310988).

\newpage
\printbibliography

\end{document}